\documentclass[pre,eqsecnum,twocolumn, showpacs]{revtex4}

\usepackage{amsmath}
\usepackage{graphicx}
\usepackage{amssymb}
\usepackage{calrsfs}

\newcommand{\kB}{k_\mathrm{B}}

\newcommand{\ris}{r_\sigma^{(i)}}
\newcommand{\rjs}{r_\sigma^{(j)}}
\newcommand{\risn}{r_{\sigma,0}^{(i)}}
\newcommand{\rjsn}{r_{\sigma,0}^{(j)}}
\newcommand{\rIs}{r_\sigma^{(1)}}
\newcommand{\rIIs}{r_\sigma^{(2)}}
\newcommand{\rIsn}{r_{\sigma,0}^{(1)}}
\newcommand{\rIIsn}{r_{\sigma,0}^{(2)}}
\newcommand{\rs}{r_\sigma}
\newcommand{\gs}{g_\sigma}

\newcommand{\Phiis}{\Phi_\sigma^{(i)}}

\newcommand{\mF}{\mathcal{F}}
\newcommand{\mFs}{\mathcal{F}_\sigma}
\newcommand{\mFsT}{{\mathcal{F}_\sigma^T}}
\newcommand{\mFsL}{{\mathcal{F}_\sigma^\mathrm{L}}}

\newcommand{\PC}{P_\mathrm{C}}
\newcommand{\FC}{\mF_\mathrm{C}}
\newcommand{\vkp}{\mathbf{k}_\perp}

\newcommand{\chii}{\chi_\sigma^{(i)}}

\newcommand{\re}{\Re\mathrm{e}}
\newcommand{\im}{\Im\mathrm{m}}
\newcommand{\msum}{\sum_{m=0}^\infty \!{}^{'}}
\newcommand{\msump}{\sum_{m,m'=0}^\infty \!\!\!\!{}^{'}}
\newcommand{\ssum}{\sum_{\sigma=s,p}}

\newcommand{\Li}{\mathrm{Li}}

\newcommand{\Tr}{\mathrm{Tr}}
\newcommand{\diag}{\mathrm{diag}}

\newcommand{\be}{\begin{equation}}
\newcommand{\ee}{\end{equation}}

\begin{document}

\title{Casimir Lifshitz pressure and free energy: exploring a simple model}
\date{\today}
\author{Simen {\AA}dn{\o}y \surname{Ellingsen}}\email{simen.a.ellingsen@ntnu.no}
\affiliation{Department of Energy and Process Engineering, Norwegian 
University of Science and Technology, N-7491 Trondheim, Norway}

\begin{abstract}
   The Casimir effect, the dispersion force attracting neutral objects to each other, may be understood in terms of multiple scattering of light between the interacting bodies. We explore the simple model in which the bodies are assumed to possess reflection coefficients independent of the energy and angle of incidence of an impinging field and show how a multitude of information can be extracted within the geometry of two parallel plates. The full thermal behaviour of the model is found and we discuss how non-analytic behaviour emerges in the combined limits of zero temperature and perfect reflection. Finally we discuss the possibility of a generalised force conjugate to the reflection coefficients of the interacting materials and how, if the materials involved were susceptible to changing their reflective properties, this would tend to enhance the Casimir attraction. The dependence of this correction on separation is studied for the constant reflection model, indicating that the effect is likely to be negligible under most experimental circumstances.
\end{abstract}

\maketitle

\section{Introduction}

The Casimir effect was first reported in 1948 \cite{casimir48} as an attractive force between parallel mirrors due to the zero point fluctuations of the electromagnetic field in vacuum. Casimir calculated the formally infinite quantum energy associated with the eigenmodes $n$ of the field between the plates, $\frac{\hbar}{2}\sum_n \omega_n$, subtracted the corresponding energy of free space (infinte plate separation) and obtained after some regularisation the simple result
\be\label{Casimir}
  P_C^0 = -\frac{\hbar c\pi^2}{240 a^4}; ~~ \mF_C^0 = -\frac{\hbar c\pi^2}{720a^3}
\ee
where $\PC$ and $\FC$ are the Casimir pressure and free energy per unit plate area respectively and $a$ is the separation between the plates. Here and henceforth a superscript $0$ refers to zero temperature. A negative pressure here corresponds to an attractive force. Naturally, the relation between pressure and free energy is $P(a) = -\partial \mF(a)/\partial a$. In the following we will employ natural units $\hbar=\kB=c=1$.

In the following section we give a brief review of the understanding of Casimir interactions as a multiple scattering or reflection phenomenon. The remainder of the paper is the beginnings of an exploration of a simple model, first employed in \cite{ellingsen08} to the author's knowledge. The model is one in which the interacting bodies scatter electromagnetic fields with reflection coefficients $|r|\leq 1$ which are modelled as invariant with respect to the energy and direction of the wave. We do not venture beyond the planar geometry herein, but show that certain closed form solutions exist in this case, and how the model enables simple extraction of key information. 

We review in section \ref{sec_closed} the derivation of closed form expressions for the Casimir force and free energy in the constant reflection model and in section \ref{sec_spectrum} how this model was used to generalise the frequency spectrum of the Casimir energy to imperfect reflection. In sections \ref{sec_thermal} through \ref{sec_ideal} we thereafter calculate the full temperature behaviour of the Casimir-Lifshitz pressure and free energy within the model and demonstrate how one encounters non-analytic behaviour in the limit of perfect reflection, reminiscent of the still ongoing debate over the temperature corrections to the Casimir force. Finally in section \ref{sec_generalised} we consider the possibility that the Casimir free energy could exhibit a generalised force on the reflective properties of the materials involved, thereby increasing its own magnitude. We lay out the basic theory of such a possibility, not hitherto reported to the author's knowledge, and use the constant reflection model to extract information about how the corresponding correction to Casimir attraction scales with temperature and separation.

\section{A brief review of the multiple scattering understanding of Casimir interactions}

The beauty and simplicity of Casimir's results (\ref{Casimir}) stems from the assumption of perfectly conducting plates, that is, the metal plates are perfect mirrors at all frequencies of the electromagnetic field. Drawing on the theory of fluctuations due to Rytov \cite{BookRytov53}, Lifshitz made an important generalisation of Casimir's results to the case of two half-spaces with frequency dependent permittivities $\epsilon_1(\omega)$ and $\epsilon_2(\omega)$ \cite{lifshitz55} (Lifshitz moreover assumed the slabs be immersed in a third medium which we assume to be vacuum here for simplicity). The calculation was rather involved and the result at zero temperature was found to be:
\begin{subequations}
\begin{align}\label{lifshitz}
  P^0 &= -\frac{1}{2\pi^2}\int_0^\infty d\zeta \int_\zeta^\infty d\kappa \kappa^2 \sum_{\sigma=s,p}\frac{\rIs \rIIs e^{-2\kappa a}}{1-\rIs \rIIs e^{-2\kappa a}} \\
  \mF^0 &= \frac{1}{4\pi^2}\int_0^\infty d\zeta \int_\zeta^\infty d\kappa \kappa \sum_{\sigma=s,p}\ln\left[1-\rIs \rIIs e^{-2\kappa a}\right]\label{lifshitzEnergy}
\end{align}
\end{subequations}
where the quantities $r^{(i)}_\sigma$ pertaining to medium $i$ are
\be\label{r}
  r^{(i)}_s = \frac{\kappa-\kappa_i}{\kappa+\kappa_i}; ~~r^{(i)}_p= \frac{\epsilon_i(i\zeta)\kappa-\kappa_i}{\epsilon_i(i\zeta)\kappa+\kappa_i}
\ee
and $\kappa_i = \kappa_i(\kappa,i\zeta) = \sqrt{\kappa^2 + [\epsilon_i(i\zeta)-1]\zeta^2}.$

By noting that $i\kappa = k_z$, $\hat z$ being the axis normal to the plates one may recognise $r^{(i)}_s$ and $r^{(i)}_p$ as the standard Fresnel reflection coefficients of a single interface for the TE and TM polarisation respectively, as well known from classical optics. Thus the Casimir-Lifshitz force (\ref{lifshitz}) does not depend directly on the bulk properties of the materials of the slabs as is ostensible from the original Lifshitz derivation, but only on the reflection properties of the \emph{surfaces} of the material half-spaces. Kats \cite{kats77} may have been the first to point this out explicitly in 1977, and the point has been given widespread attemtion more recently \cite{jaekel91, lambrecht97, genet03, lambrecht06}. It is a simple exercise to show that inserting $(r^{(i)}_\sigma)^2=1, ~~\forall i, \sigma$ into (\ref{lifshitz}) and (\ref{lifshitzEnergy}) yields the Casimir limits (\ref{Casimir}).

The trait that the Casimir-Lifshitz pressure (\ref{lifshitz}) is a function of reflection properties only is a tell-tale that the effect may be thought of as the result of multiple scattering of light between boundaries. Another hint is the recognition of the fraction in (\ref{lifshitz})
\be
  \frac{\rIs \rIIs e^{-2\kappa a}}{1-\rIs \rIIs e^{-2\kappa a}} = \sum_{k=1}^\infty \left(\rIs \rIIs e^{2ik_z a}\right)^k
\ee
as a sum of contributions from waves which are reflected off both interfaces $k$ times before returning to whence it originated.

This implies that the Casimir interaction between much more general materials than bulk dielectrics (as considered by Lifshitz) may be calculated, if one is able to obtain an expression for the reflection properties of the surfaces involved and how light is transmitted between the bodies. This fact was used, among other things, to calculate the effect of spatial dispersion \cite{kats77,esquivel03,svetovoy05, sernelius05} and interaction between (magneto)dielectric multilayers \cite{tomas02,raabe03,tomas05, henkel05,ellingsen07} based on Green's function methods \cite{tomas95}. Some further considerations were given in \cite{ellingsen07b}.

In recent years, the understanding of Casimir problems in terms of multiple scattering has become widespread and makes way for what is presently perhaps the most powerful techniques for calculating Casimir energies in non-trivial geometries. Within such a general scattering formalism the Lifshitz formula (\ref{lifshitzEnergy}) may be seen as a special case of the much more general formula
\be\label{TGTG}
  \mF^0 = \int_0^\infty \frac{d\zeta}{2\pi} \Tr \ln \left[1-\mathbb{T}_1\mathbb{G}^0_{12}\mathbb{T}_2\mathbb{G}^0_{21}\right]
\ee
where $\mathbb{T}_i$ is the T-matrices (operators) of two arbitrary interacting bodies and $\mathbb{G}^0_{ij}$ is a vacuum propagator (Green's function) from object $i$ to object $j$. The energy expression (\ref{TGTG}) was recently dubbed the TGTG formula and is written here as derived in \cite{kenneth06, kenneth07}, but the use of less general embodiments of essentially the same multiple scattering technique goes back at least to the 1970s \cite{renne71, balian78}. The recent acceleration of progress towards understanding the role of geometry in Casimir interactions has brought much attention to this technique in recent years (e.g.\ \cite{bordag06,emig07, milton08, milton08b, milton08d}; for a review see \cite{milton08c} and the introduction to \cite{milton08b}).

To see somewhat roughly how the Casimir-Lifshitz free energy (\ref{lifshitzEnergy}) is a special case of (\ref{TGTG}) let the propagators be simply that of a plane wave along the $\hat{z}$ direction over a distance $a$, $\mathbb{G}^0\to \exp(ik_z a)$ and let the T matrices represent specular scattering at the surfaces, $\mathbb{T}_i \to \diag(r_s^{(i)}, r_p^{(i)})$. Take the trace operation in (\ref{TGTG}) to include an integral over the transverse momentum $\mathbf{k}_\perp$ plane (isotropic due to rotational symmetry) and one obtains (\ref{lifshitzEnergy}) with minimal manipulation. See e.g.\ \cite{milton04} for details.
 
For reasons of simplicity much of the recent research on geometry effects has been made for the massless scalar field satisfying the Klein-Gordon equation rather than the vectorial electromagnetic field. Historically, Dirichlet and Neumann boundary conditions have been employed together with path integral methods of quantum field theory to mimic the two electromagnetic polarisations (note that the sum of the Dirichlet and Neumann scalar solutions of the wave equation only reproduces the ideally conducting electromagnetic case in special geometries where the electromagnetic modes decouple, such as the original Casimir geometry). 

In order to model semi-transparent bodies in this formalism, the introduction of delta-function potentials into the Klein-Gordon equation has been common (see review in \cite{milton04}). A delta potential $V(\mathbf{r}) = \lambda \delta^3(f(\mathbf{r}))$ models a body whose surface solves $f(\mathbf{r})=0$ and where the coupling constant $\lambda$ determines the ``transparency''. Dirichlet boundary conditions are regained in the strong coupling limit $\lambda\to \infty$, and it has turned out that several non-trivial geometries are exactly solvable to linear order in $\lambda$ in the weak coupling case $\lambda\ll 1$ \cite{milton08, milton08b}. 

The model of constant reflection coefficients is a somewhat similar idea and constitutes another model of semi-transparency where some physicality is traded for mathematical manageability.

\section{Closed form expression using polylogarithms}\label{sec_closed}

It is straightforward to obtain a closed form expression for the Casimir pressure and energy in the constant reflection model. The mathematical formalism which enters is that of polylogarithmic functions. The $\nu$th order polylogarithm of $x$ is defined as
\be\label{polylog}
  \Li_\nu(x) = \sum_{k=1}^\infty \frac{x^k}{k^\nu}.
\ee
It is related to the Riemann zeta functions (as is obvious for $\nu > 1$) by $\Li_\nu(1)=\zeta(\nu)$ and obeys the recursion relation $(d/dx) \Li_\nu(x) = (1/x)\Li_{\nu-1}(x)$, which in particular implies that for $|\re A|<1$
\be\label{Liint}
  \int dx \Li_\nu(A e^{-bx})= -\frac{1}{b}\Li_{\nu+1}(A e^{-b x}) + C
\ee
where $A,b,C$ are constants. We recognise the polylogarithms which enter into (\ref{lifshitz}) and (\ref{lifshitzEnergy}),
\be
  \Li_1(x) = -\ln(1-x); ~~~\Li_0(x) = \frac{x}{1-x}.
\ee
The polylogarithms of interest herein are all of real and integer order.

\begin{figure}[tb]
  \includegraphics[width=2in]{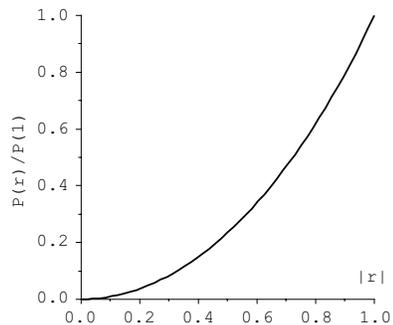}
  \caption{Casimir pressure as a function of a constant reflection coefficient relative to the ideal conductor Casimir result. Materials are assumed similar and the reflection coefficient equal for both polarisations for simplicity.}\label{fig_Pr}
\end{figure}

In the Wick rotated formalism in Euclidean space where the time axis is imaginary, it follows from the general properties of causal response functions that the reflection coefficients are necessarily real quantities \cite{BookLandau80}. Now, assuming the reflection coefficients are constants with respect to $\kappa$ and $\zeta$ the integrals are easily solved with partial integration using (\ref{Liint}) and yields for the pressure and free energy at zero temperature, respectively\footnote{If the calculation is performed for real frequencies, reflection coefficients are generally complex and the real part of the $\Li_4$ functions should be taken \cite{ellingsen08}.},
\begin{align}
  P^0 &= -\frac{3}{16\pi^2 a^4} \sum_{\sigma=p,s} \Li_4(\rIs\rIIs);\label{P0}\\
  \mF^0 &= -\frac{1}{16\pi^2 a^3} \sum_{\sigma=p,s} \Li_4(\rIs\rIIs)\label{F0}.
\end{align}
In the ideal limit $|\rs|\to 1$, $\Li_4(\rIs\rIIs)\to \zeta(4)=\pi^4/90$ and Casimir's results (\ref{Casimir}) are regained. The Casimir pressure as a function of the squared reflection coefficient $r^2$ (assuming both materials equal and the same coefficient for both polarisations) is plotted in figure \ref{fig_Pr}. A similar graph for the free energy would obviously be exactly identical.

\section{Real-frequency spectrum}\label{sec_spectrum}

The model of constant reflections was introduced in \cite{ellingsen08} in order to slightly generalise considerations of the real-frequency spectrum of the Casimir force due to Ford \cite{ford93}. He showed from quantisation of the vacuum how the Lifshitz frequency integrand is equal to the vacuum energy spectrum, which in the case of perfect mirrors studied by Ford turns out to be an oscillating function of frequency with discontinuities at $\omega = n \pi/a,~ n\in \mathbb{N}$. The Lifshitz pressure formula for real frequencies at zero temperatures reads \cite{lifshitz55}
\begin{align}
  P^0(a) = -\frac{1}{2\pi^2 }\re &\int_0^\infty  d \omega\omega^3 \int_\Gamma  d p p^2 
  \notag\\
  &\times\sum_{\sigma=s,p} \frac{r^2_\sigma \exp(2ip\omega a)}{1-r^2_\sigma \exp(2ip\omega a)}\label{realLifshitz}
\end{align}
where the Lifshitz variable $p$ is the positive real part root of $p = \sqrt{1-(\mathbf{k}_\perp/\omega)^2}$. In the following we will assume the materials equal for simplicity; the generalisation to different reflectivity is $\rs^2\to\rIs\rIIs$. Replacing an isotropic integral over all $\mathbf{k}_\perp$  the integration contour $\Gamma$ therefore runs from $1$ to $0$ (propagating modes) and thence to $i\infty$ (evanescent modes). 

\begin{figure}[tb]
  \includegraphics[width=2.5in]{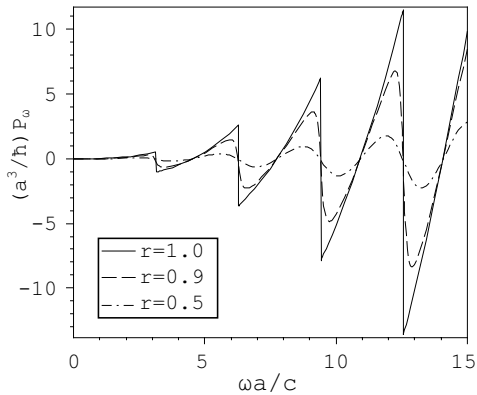}
  \caption{Casimir-Lifshitz frequency spectrum for real constant reflection coefficients. This figure generalises figure 2 of \cite{ford93}.}\label{fig_graph}
\end{figure}

By assuming reflection coefficients to be constant with $|\re\{\rs^2\}|\leq 1$, the frequency spectrum can be found. Defining
\be
  P^0 = \int_0^\infty  d \omega \ssum P^0_{\omega,\sigma}
\ee
one finds the spectrum
\begin{align}
   P^0_{\omega,\sigma} =&\frac{-1}{16\pi^2a^3} \left[-\xi^2\im\Li_1(\rs^2 e^{ i\xi})\right.\notag\\
   &\left.-2\xi\re\Li_2(\rs^2 e^{ i\xi})+2\im\Li_3(\rs^2 e^{ i\xi})\right]\label{spectrum}
\end{align}
where we have defined the shorthand dimensionless quantity $\xi = 2\omega a$. The spectrum (\ref{spectrum}) is plotted for a few different $\rs$ in figure \ref{fig_graph}. Note how the discontinuous behaviour seen in the ideal case $\rs^2=1$, which stems from the term
\be
  \im\Li_1(e^{i\xi}) = \arctan\left(\frac{\sin\xi}{1-\cos\xi}\right)
\ee
becomes smooth for $\rs^2 < 1$. This is one example of how the Lifshitz formulae exhibit non-analytic behaviour in the perfectly reflecting limit, a fact which is closely related to the ongoing dispute about the temperature correction to the Casimir force as explained in the following.

\section{Thermal behaviour}\label{sec_thermal}

We start by generalising the closed form result (\ref{F0}) to include finite temperature corrections. It is easiest to work within the imaginary frequency formalism. When going to finite temperature the real frequency integrand of (\ref{realLifshitz}) and the corresponding free energy expression receives an additional factor $\coth(\omega/2T)$ from the Bose-Einstein distribution. By use of Cauchy's theorem the real frequency integral can be written as a sum over the poles of this factor at $\omega/2T = m\pi i,~m\in \mathbb{N}$. Thus the Lifshitz formula for free energy of polarisation mode $\sigma$ (letting $\mF = \mF_p + \mF_s$) at temperature $T$ is
\begin{subequations}
\begin{align}
  \mFsT =& \frac{T}{2\pi}\msum \int_{\zeta_m}^\infty  d \kappa\kappa \ln(1-\rs^2 e^{-2\kappa a}) \label{FT}\\
    =& 
    -\frac{T}{8\pi a^2}\msum \left[2a\zeta_m\Li_2(\rs^2 e^{-2\zeta_m a})\right.\notag\\
    &+\left.\Li_3(\rs^2 e^{-2\zeta_m a})\right]\label{Fintermediate}
\end{align}
\end{subequations}
where $\zeta_m = 2\pi m T$ are the Matsubara frequencies and the prime on the sum means the $m=0$ term is taken with half weight. In the last form we use that $\ln(1-x)=-\Li_1(x)$, and partial integration by use of (\ref{Liint}). 

In the high temperature limit $2\zeta_1 a \gg 1$ the $m=0$ term dominates (other terms are exponentially small) and we immediately obtain the free energy in this limit:
\be  \label{hiTF}
  \mFsT \sim -\frac{T}{16\pi a^2}\Li_3(\rs^2);~~ \zeta_1 a \gg 1,
\ee
in accordance with the well known high-temperature free energy between ideal plates, $\mF_C \approx -\zeta(3)T/(8\pi a^2)$ known at least since the 1960s \cite{mehra67}. 

By using the definition (\ref{polylog}) and changing the order of summation, (\ref{Fintermediate}) can be written
\be\label{FTinterm}
  \mFsT = \frac{-T}{16\pi a^2} \sum_{k=1}^\infty\frac{\rs^{2k}}{k^3}\left[\frac{\zeta_ka}{\sinh^2(\zeta_ka)}+\coth(\zeta_ka)\right]
\ee
This is a generalisation of equation (3.12) of \cite{hoye03}, which is for ideal conductors. One may note that the expression between the square brackets equals the Wronskian $\mathcal{W}(\coth x, x)$ with $x=\zeta_k a$. For numerical purposes (\ref{FTinterm}) is useful for having a summand which converges geometrically and consists of standard functions only.

\begin{figure}[tb]
  \includegraphics[width=3in]{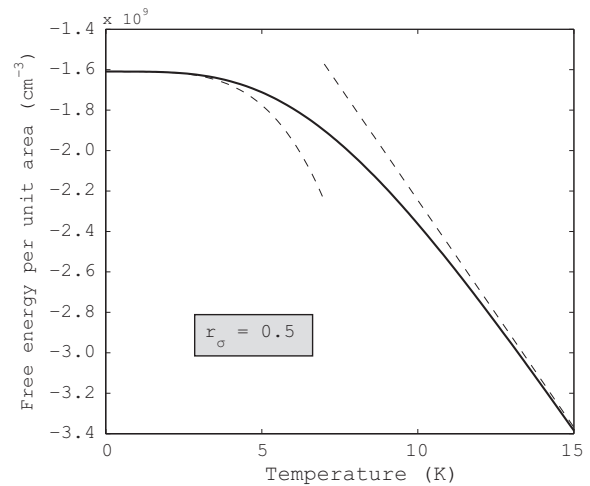}
  \caption{Casimir-Lifshitz free energy as a function of temperature for $\rs=1/2$ and the high and low temperature asymptotics, (\ref{hiTF}) and (\ref{lowTF}) respectively.}\label{fig_FT}
\end{figure}

We go on to find the asymptotic behaviour for small $T$. When $aT$ is small and $\rs^2<1$ only small values of the quantity $\zeta_k a$ are of importance to the sum (\ref{FTinterm}) because for a given $\rs$ the temperature may be chosen so small that the sum has converged due to the factor $\rs^{2k}$ before $\zeta_k a$ becomes of order unity. Then a Laurent expansion 
\[
  x\sinh^{-2}(x)+\coth(x) = 2x^{-1} + 2x^3/45+...
\]
gives the low temperature expansion assuming $\rs^2<1$:
\begin{align}
  \mFsT \sim& -\frac{1}{16\pi^2a^3}\Li_4(\rs^2) - \frac{\pi^2 a T^4}{45}\frac{\rs^2}{1-\rs^2}\notag\\
  &+ \mathcal{O}(T^6); ~~~T\to 0\label{lowTF}
\end{align}
where we use $\Li_{0}(x)=x/(1-x)$. The thermal behaviour of $\mFs$ is plotted in figure \ref{fig_FT} together with the high and low temperature asymptotics.

One may note a couple of peculiar traits about this low-temperature behaviour. Firstly, all finite temperature coefficients are singular in the ideal limit $\rs^2\to 1$; there are only even order terms, and the temperature correction of order $T^{2n}$ diverges as $(1-\rs^2)^{3-n}$ for $n\geq 2$ as we will show below. This is an indication that $\mFsT$ is not analytic in the double limit where $T$ vanishes and $\rs^2\to 1$. 

Secondly, note the contrast with the corresponding ideal result $\rs^2=1$ derived in \cite{hoye03, brown69},
\be\label{FTCasimir}
  \frac1{2}\mF_{C}^T \sim -\frac{\pi^2}{1440 a^3} - \frac{\zeta(3)T^3}{4\pi} + \frac{\pi^2 a T^4}{90} +...; ~~T\to 0.
\ee
where further corrections are exponentially small (see also \cite{BookMilton01}). Mathematically the change of sign and coefficient of the $T^4$ term from (\ref{lowTF}) to (\ref{FTCasimir}) can be na\"{i}vely explained by
\be\label{zetareg}
  \frac{\rs^2}{1-\rs^2} = \Li_0( \rs^{2}) \buildrel{\rs^2\to 1}\over{\longrightarrow} \zeta(0) = -\frac1{2},
\ee
yet there appears a hitherto unseen term $\propto T^3$ which is independent of $a$ and therefore does not contribute to the Casimir pressure. 

Mathematically, the reason for this fundamental change of temperature behaviour at $\rs^2=1$ is due to the fact that the summand of (\ref{FT}) becomes a non-analytical function of $m$ at $m=0$ when $\rs^2=1$, but is analytical whenever $\rs^2<1$. It was demonstrated in \cite{ellingsen08b} that a term $\propto T^3$ in the low temperature expansion of $\mF$ appears when the summand of (\ref{FT}) contains a term proportional to $m^2\ln(m)$.

Before elaborating this further, we will work out the full asymptotic series expansion of $\mF$ in powers of $T$ by use of the method developed in \cite{ellingsen08b}. We define the function $\gs(\mu)$
\be\label{gmu}
  \mFsT \equiv -\frac{T}{8\pi a^2} \msum \gs(\mu)
\ee
where $\mu=mT$ and $\gs(\mu)$ is the expression inside the square brackets of (\ref{Fintermediate}). When $\gs(\mu)$ is analytical at $\mu=0$, $\gs$ can be written as a Taylor series $\gs(\mu) = \sum_{k=0}^\infty c^\sigma_k \mu^k$. By zeta regularisation the temperature correction $\Delta \mFs(T) = \mFsT-\mFs^0$ can be written\cite{ellingsen08b}
\begin{align}
  \Delta\mFs(T) &\sim -\frac{1}{8\pi a^2} \sum_{k=1}^\infty c^\sigma_{2k-1}\zeta(1-2k)T^{2k}\notag\\ 
  &= \frac{1}{8\pi a^2} \sum_{k=1}^\infty c^\sigma_{2k-1}\frac{B_{2k}}{2k}T^{2k}; ~~~T\to 0,\label{DFT}
\end{align}
where $B_n$ are the Bernoulli numbers as defined in \cite{BookAbramowitz64}. Only odd orders of $\mu$ from the Taylor expansion contribute since $\zeta(-2k)=0$; $k\in \mathbb{N}$, thus there are only even orders of $T$.

Since
\[
 \left(\frac{d}{dx}\right)^k \Li_n(Ae^{-bx}) = (-b)^k\Li_{n-k}(Ae^{-bx})
\]
and since for $\re A<1$,
\[
  \Li_{-k}(A) \propto (1-A)^{-(k+1)},~~k\geq 0,
\]
it is clear that the summand of (\ref{FT}) is analytic if and only if $\rs^2<1$, since the higher derivatives of the $\Li_3$ term become divergent at $m=0$. The asymptotic series on the form (\ref{DFT}) is therefore valid for all $\rs^2<1$ but not in the perfectly reflecting limit.

When $\rs^2<1$ it is obvious that
\[
  \Li_n (\rs^2  e^{-\alpha}) = \sum_{l=0}^\infty \frac{(-\alpha)^l}{l!} \Li_{n-l}(\rs^2),
\]
which automatically gives the Taylor expansion of $\gs(\mu)$. Inserted into $\gs(\mu)$ from (\ref{Fintermediate}) we find
\begin{align}
  \gs(\mu) =& \Li_3(\rs^2) 
  - \sum_{k=1}^\infty \frac{k-1}{k!}(-4\pi a\mu)^k\Li_{3-k}(\rs^2).\label{gsmur}
\end{align}
It is thus clear that $c^\sigma_1=0$, in accordance with (\ref{lowTF}) where the lowest correction to zero temperature was found to be $T^4$. With (\ref{DFT}) the full temperature expansion to arbitrary order is thus
\begin{align}
  \mFsT 
=& \frac{1}{16\pi^2a^3}\sum_{k=0}^\infty \frac{(k-1)B_{2k}}{(2k)!}\Li_{4-2k}(\rs^2) (4\pi a T)^{2k}.\label{FTasymptotic} 
\end{align}
One may easily verify that this generalises (\ref{lowTF}), noting that $\Li_0(x) = x/(1-x)$. One may show that this series has zero convergence radius, that is, it does not converge for any finite $T$.

\section{Asymptotic temperature expansion for perfect conductors revisited}\label{sec_ideal}

The fact that the na\"{i}ve transition (\ref{zetareg}) yields the correct $T^4$ term for ideal conductors leads one to speculate that the even-power terms of the asymptotic $T$-series for ideal conductors may be given by simply letting $\Li_{4-2k}(\rs^2)\to \zeta(4-2k)$ in (\ref{FTasymptotic}). Since the Riemann zeta function with even negative integer arguments is zero, this would if so truncate the series beyond order $T^4$. This does not explain the appearence of the $T^3$ term in (\ref{FTCasimir}), however, and does not preclude the emergence of other additional terms of higher non-even order.

The answer is readily found using the above mentioned method developed in \cite{ellingsen08b}. From (\ref{Fintermediate}) and (\ref{gmu}) we see that for ideal conductors
\be
  \gs(\mu) = \tau\Li_2(e^{-\tau}) + \Li_3(e^{-\tau})
\ee
where we have defined the shorthand $\tau = 4\pi a \mu$. The asymptotic behaviour of $\Li_n(e^{-\tau})$ for small $\tau$ was found by Robinson \cite{robinson51} who studied the function\footnote{For this integral representation of the polylogarithm see e.g.\ \cite{kolbig70} equation (2.4).}
\[
  \phi(s, \tau) = \frac1{\Gamma(s)}\int_0^\infty dx \frac{x^{s-1}}{e^{x+\tau}-1} = \Li_{s}(e^{-\tau}).
\]
For integer $s=n$ the Robinson formula is
\begin{align}
  \Li_n(e^{-\tau}) =& \frac{(-\tau)^{n-1}}{(n-1)!}\left[\sum_{k=1}^{n-1}\frac1{k}-\ln(\tau)\right] \notag \\
  &+ \sum_{\mbox{\tiny$\begin{array}{c} k=0 \\ k\neq n-1 \end{array}$}}^{\infty} \frac{\zeta(n-k)}{k!}(-\tau)^k\label{Robinson}
\end{align}
which gives
\begin{align}
  \gs(\mu) =& \zeta(3)-\frac{\tau^2}{4}+\frac1{2}\tau^2\ln(\tau)\notag\\
  &-\sum_{k=3}^\infty\frac{k-1}{k!}\zeta(3-k)(-\tau)^k.\label{gmuFull}
\end{align}
It is shown in \cite{ellingsen08b} that, as defined in (\ref{gmu}), a term in $\gs(\mu)$ of the form $c_{2l}^\sigma \mu^2 \ln\mu$ gives a term in the free energy
\[
  \mF_{2l,\sigma} = -\frac{1}{8\pi a^2}\frac{\zeta(3)}{4\pi^2}c_{2l}^\sigma T^3.
\]
From (\ref{gmuFull}) one recognises $c_{2l}^\sigma = 8\pi^2 a^2$, wherewith the $T^3$ term of (\ref{FTCasimir}) is regained. 

Terms of $\gs(\mu)$ which are constant or proportional to $\mu^2$ give no contribution to the temperature correction to free energy and a comparison of (\ref{gsmur}) and (\ref{gmuFull}) to order $\mu^3$ and higher shows that for all orders of $T$ above cubic the expansion of $\mF^T_C$ is the same as (\ref{FTasymptotic}) with $\Li_{2-2k}(\rs^2)\to \zeta(2-2k)=-\frac{1}{2},0,0,...$ for $k=1,2,3,..$. Thus the series is terminated at fourth order and the expansion (\ref{FTCasimir}) is in fact the full temperature behaviour modulo exponentially small corrections:
\be
  \mF_{C}^T \sim -\frac{\pi^2}{720 a^3} - \frac{\zeta(3)T^3}{2\pi} + \frac{\pi^2 a T^4}{45}; T\to 0.
\ee
This result was found by different methods in \cite{hoye03,brown69, BookMilton01} and is consistent with Mehra's early considerations \cite{mehra67}. 

\subsection{Relation to the temperature debate}

In connection with an ongoing debate concerning the temperature correction to the Casimir force, a point which has been raised is that the application of certain reflectivity models lead to apparent inconsistencies with the third law of thermodynamics, the Nernst heat theorem (c.f.\ \cite{klimchitskaya06} and references therein), that is, entropy does not vanish with vanishing temperature as it should. It was recently concluded that these formal violations of Nernst's theorem stem from non-analytical behaviour in the combined limit of zero frequency (where reflection coefficients approach unity for metal models) and zero temperature \cite{intravaia08, ellingsen08c}. Indeed, violation can only occur due to particular types of non-analyticities causing abrupt change of reflectivity at the point $\omega=T=0$ \cite{ellingsen08d}. The nonzero entropy at zero temperature would then stem from the fact that the summand of the free energy sum such as (\ref{FT}) became discontinuous at $m=0$.

The transition from imperfect to perfect reflection in the previous paragraph is reminiscent of the anomalous entropy at some level. In \cite{intravaia08,ellingsen08c,ellingsen08d} the situation is one in which the reflection coefficients and thus the free energy summand is discontinuous when frequency and temperature are taken continuously to zero. Here the second temperature \emph{derivative} of the free energy integrand (\ref{FT}) is discontinuous (indeed divergent) as reflection coefficient and temperature are taken continuously to zero. The former discontinuity leads to a change in free energy leading temperature dependence from quadratic to linear, the linear dependence which implies nonzero entropy at zero temperature since $S=-\partial\mF/\partial T$. The $\rs\to 1$ transition considered above changes the temperature correction from quartic to cubic. No anomalous entropy at $T=0$ stems from this transition, yet its mathematical dynamics are very similar. 

\section{A generalised force on reflectivity?}\label{sec_generalised}

We conclude with a few remarks on the possibility of a generalised force whose generalised coordinate is the reflectivity of one of the materials. In most calculations of Casimir forces between real materials the material is treated as inert and it is assumed that its reflection properties do not change due to the Casimir interaction across the gap. One could remark, however, that were it possible, the system could lower its free energy by increasing its reflectivity. Such a mechanism was in fact suggested as a possible explanation of the energetics of the high temperature superconducting transition in which a ceramic multilayer can decrease its total free energy by becoming superconducting, thus a better reflector \cite{kempf08}. 

In the following a few notes are made on this possibility. A determination of the question of whether such an effect could be measurable is only possible subsequent to calculating the material's free energy as a functional of its reflection coefficients and determining to which extent variation of reflectivity is a degree of freedom. This is complicated task we shall not pursue herein.

One is reminded at this point of the previously mentioned dispute over the thermal dependence of the Casimir effect between real materials (reviews include \cite{brevik06, klimchitskaya06}). Puzzlingly, recent high accuracy experiments which have measured the Casimir force between good metals (\cite{bezerra06} and references therein) report a measured Casimir pressure significantly larger than that predicted by several theoretical groups \cite{hoye03,brevik06,bostrom00,buenzli08}. 

Our calculations indicate that the Casimir self-enhancing effect is negligible under most circumstances yet it might be worth investigating it further taking into account specific material characteristics for a quantitative treatment. Here we shall content ourselves with laying out the very basic theory and using the constant reflection model as a tool to extract the dependence on temperature and separation in two limits.

Consider the Lifshitz free energy on yet another form, 
\begin{align}
  \mFsT[\rIs, \rIIs] =& \frac1{2i} \int_{-\infty}^\infty \frac{d\omega}{2\pi}\coth \frac{\omega}{2T} \notag \\
  &\times \int \frac{d^2 k_\perp}{(2\pi)^2} \ln(1-\rIs\rIIs e^{-2\kappa a})\label{FTomkp}
\end{align}
with $\kappa = \sqrt{\vkp^2-\omega^2}$ with $\re\{\kappa\}>0$ and reflection coefficients functions of $\vkp$ and $\zeta$. In the special case of a single interface between vacuum and a dielectric, $r^{(i)}_\sigma$ take the form (\ref{r}). Note that the integrand of (\ref{FTomkp}) is complex but only the imaginary part contributes due to symmetry properties so that the expression as a whole is real (see e.g. \cite{ellingsen08c}). The logarithm is understood as its principal value.

The total free energy of the system per unit transverse area should be well approximated by 
\[
  \mFs^\text{tot} = \mFs^{(1)}[\rIs] + \mFs^{(2)}[\rIIs] + \mFs^\text{L}[\rIs,\rIIs]
\]
where the first two terms on the right hand side pertain to the two media on either side of the gap and the last term is the Lifsthiz free energy, now with a superscript L for distinction (we assume finite temperature throughout this section except as explicated). We define the generalised force acting on material $i$:
\begin{align}
  \Phiis(\omega, \vkp) =& - \frac{\delta \mFs^\text{L}[\ris, \rjs]}{\delta \ris(\vkp,\omega)} \notag \\
  =& \frac1{2i}\coth\left(\frac{\omega}{2T}\right)\frac{\rjs e^{-2\kappa a}}{1-\ris\rjs e^{-2\kappa a}}
\end{align}
where $i,j=1,2$; $i\neq j$ and $\delta/\delta \ris$ denotes the functional derivative. The dependence of reflection coefficients on $\omega$ and $\vkp$ has been suppressed on the right hand side. 
The generalised force can take either sign but always acts so as to increase the attraction between the plates, an observation which is self evident from the fact that the negative Casimir-Lifshitz free energy (\ref{FT}) increases in magnitude with increasing reflectivity \footnote{One may note that if one were to have a dielectric and a magnetic material, repulsion can in principle be effectuated. In this case $\Phi_\sigma$ acts to decrease repulsion.}. 

A given material $i$ will have a generalised susceptibility which determines its ability to alter its reflective properties in response to $\Phi_\sigma$,
\begin{align}\label{chir}
  \chii(\omega,\omega',&\vkp,\vkp') = \frac{\delta \ris(\omega,\vkp)}{\delta\Phiis(\omega',\vkp')}\\
  &= \left[\frac{\delta^2 \mFs^{(i)}[\ris]}{\delta \ris(\omega,\vkp)\delta \ris(\omega',\vkp')}\right]^{-1}\label{Fdouble}
\end{align}
and a Taylor expansion in $\Phi_\sigma$ gives
\begin{align*}
  \Delta \ris(\omega,\vkp) =& \int_{-\infty}^\infty\frac{d\omega'}{2\pi}\int \frac{d^2k'_\perp}{(2\pi)^2}\chii(\omega,\omega',\vkp,\vkp')\\
  &\times\Phiis(\omega',\vkp')+...
\end{align*}
At finite temperature we may close the $\omega'$ integral path around the upper half complex plane and invoke the Cauchy theorem. Since $\chii(\cdots)$ does not have any singularities in the upper $\omega'$ plane \cite{BookLandau80}, the integral over $\omega'$ then gives a sum over the poles of $\coth(\omega'/2T)$, and by letting $\omega\to i\zeta$ we obtain 
\begin{align}
  \Delta \ris(i\zeta,\vkp) = T\msum &\int \frac{d^2k'_\perp}{(2\pi)^2}\chii(i\zeta,i\zeta_m,\vkp,\vkp')\notag\\
  &\times\Phiis(i\zeta_m,\vkp')+...\label{deltar}
\end{align}
where
\be\label{Phi}
  \Phi^{(i)}_\sigma(i\zeta,\vkp) = \frac{r^{(j)}_\sigma(i\zeta,\vkp) e^{-2\kappa a}}{1-r^{(i)}_\sigma(i\zeta,\vkp) r^{(j)}_\sigma(i\zeta,\vkp) e^{-2\kappa a}}.
\ee
On the imaginary frequency axis all quantities in (\ref{deltar}) and (\ref{Phi}) are real.

Since $\chii(\cdots)$ depends on $\ris$ and $\Phiis$ depends on both reflection coefficients, quation (\ref{deltar}) defines a set of integral equations for the new reflection coefficients. Note that $\Phiis$ always has the same sign as $\ris$ and increases in magnitude with increasing $|\ris|$, so equation (\ref{deltar}) implies that given time, $|\ris|$ will flow to ever higher values until the fixed point
\be\label{fixedpoint}
  \chii(i\zeta,i\zeta,\vkp,\vkp) = 0
\ee
is reached for both materials. If one is able to calculate $\chii(\cdots)$ for a given $\ris$, (\ref{deltar}) with (\ref{Phi}) may be invoked iteratively for a simple numerical scheme to obtain the new reflection coefficients. 

An approximation of the change in reflectivity is provided by use of (\ref{deltar}) using the 'first order' estimate 
\be\label{Phin}
  \Phi_{\sigma,0}^{(i)} = \frac{\rjsn e^{-2\kappa a}}{1-\risn\rjsn e^{-2\kappa a}}
\ee
where $\risn$ are the reflection coefficients without any Casimir interaction, which satisfy $\delta \mFs^{(i)}/\delta \risn = 0$. To first order in $\Delta r$ the change in Lifshitz free energy is
  \begin{align}
    \Delta\mFsL =& -T\msum \int \frac{d^2 k_\perp}{(2\pi)^2} \left(\frac{\Delta \rIs}{\rIsn}+ \frac{\Delta\rIIs}{\rIIsn}\right)\notag \\ 
    &\times\frac{\rIsn\rIIsn e^{-2\kappa a}}{1-\rIsn\rIIsn e^{-2\kappa a}}.
  \end{align}
which, upon comparison with (\ref{Phin}) gives the 'one-loop' approximation
\begin{widetext}
  \be\label{one-loop}
    \Delta\mFsL \approx -T^2\msump \int \frac{d^2 k_\perp}{(2\pi)^2}\frac{d^2 k'_\perp}{(2\pi)^2}\sum_{i=1,2} \Phi_{\sigma,0}^{(i)}(i\zeta_m, \vkp)\chii(i\zeta_m,i\zeta_{m'},\vkp,\vkp') \Phi_{\sigma,0}^{(i)}(i\zeta_{m'}, \vkp').
  \ee
\end{widetext}
It is understood that $\chii(\cdots)$ is evaluated assuming unperturbed reflection.


\subsection{Constant reflection model}

Assuming constant reflection coefficients as before it is easy to see that $\Phi_\sigma$ scales with distance like $\mFs$:
\begin{align}
  \Phi^{(i)}_\sigma &= T\msum \int \frac{d^2k'_\perp}{(2\pi)^2}\Phi^{(i)}_\sigma(i\zeta_m,\vkp) =-\frac{\partial \mFs}{\partial \ris} \notag \\ 
  &\sim \left\{\begin{array}{cc} (16\pi^2 a^3 \ris)^{-1} \Li_3(\rIs\rIIs), & T\to 0 \\ T(16\pi a^2\ris)^{-1}\Li_2(\rIs\rIIs), & aT\gg 1\end{array}\right.\label{Phiconst}
\end{align}
where only the last form is specific to the constant reflection model. 

The one-loop correction (\ref{one-loop}) now simplifies to 
\begin{align*}
  \Delta\mFsL \approx& -\frac{T^2}{4\pi^2}\msump \int_{\zeta_m}^\infty d\kappa \kappa\int_{\zeta_{m'}}^\infty d\kappa' \kappa' \\
  &\times\sum_{i=1,2} \Phi_{\sigma,0}^{(i)}(\kappa) \chii(\kappa,\kappa')\Phi_{\sigma,0}^{(i)}(\kappa').
\end{align*}
The dependence of $\chii$ on $\kappa,\kappa'$ is of course unknown, but it is in the spirit of our simple model to assume it constant with respect to these arguments (dependent on $\rIs$ and $\rIIs$ only) as a first approximation so as to extract some information as to how the corrections to Casimir force and free energy depend on distance. In this model the simple result is
\be
  \Delta \mFsL \approx -\chii \left[\Phiis\right]^2\propto \left\{\begin{array}{cc} a^{-6}, & T=0 \\ T^2a^{-4}, & aT\gg 1\end{array}\right.
\ee
with $\Phiis$ from (\ref{Phiconst}).

The indication is thus that the change in the Casimir pressure will fall off as $a^{-7}$ and $a^{-5}$ in the two regimes respectively, much faster than the Casimir pressure, which falls off as $a^{-4}$ and $a^{-3}$ respectively. Although tentative and subject to restrictive assumptions, the above calculation indicates that the effect of the generalised force on reflectivity is likely to be negligible under most circumstances. It is notable, however, that the effect increases as $T^2$ in the high $aT$ limit, whereas the Casimir force is a linear function of temperature in this regime. 

\section*{Conclusions}

We have reviewed how the Casimir effect can be thought of as a multiple scattering phenomenon, an observation which inspires the use of a simple model in which the reflection coefficients of interacting bodies (the relative amplitude of reflected vs.\ incoming field) are assumed to be independent of the direction and energy of the field. We review how this simple model yields some closed form results in the planar geometry famoyusly considered by Casimir and Lifshitz, and how much important information may be extracted with relatively simple methods within the confines of the model. 

We review how the frequency spectrum of the Casimir effect is generalised from perfect reflection and becomes analytic and continuous upon introducing non-unity reflection coefficients. The full asymptotic behaviour of the Casimir-Lifshitz free energy in powers of temperature is found, and it is demonstrated how the transition to the perfectly reflecting case is not smooth. This is another demonstration of the non-analytic behaviour of the Lifshitz formalism in the double limit of zero temperature and perfect reflection which has given rise to debate over the thermodynamic consistency of various reflection models in connection with the temperature behaviour of the Casimir force. 

We finally discuss the idea of a generalised ``Casimir'' force conjugate to the reflection coefficients of the interacting bodies. If there exist mechanisms by which the materials involved could be susceptible to changing their reflective properties, the generalised force initiates a back reaction effect by which the reflection coefficients tend towards their maximal available values, increasing the Casimir interaction. The indication is, however, that the effect would be small and fall off faster with interplate separation than the Casimir force itself.

\section*{Acknowledgements}

I wish to express my gratitude to Professor Iver Brevik who opened my eyes to the Casimir universe, whose formidable insight into many and diverse areas of physics is a constant source inspiration, and without whose careful supervision and assiduous support I could not have come this far. Further thanks go to Professor Kimball A.\ Milton for many useful comments on this manuscript.

\end{document}